%% file: sc.tex
\documentclass[twocolumn, prl,secnumarabic,  floatfix, nofootinbib, nobalancelastpage, longbibliography,aps,10pt]{revtex4-2}
\usepackage{xr}
\usepackage{amsmath, graphicx, array, amssymb}
\usepackage[usenames,dvipsnames]{color}
\usepackage[colorlinks=true,allcolors=Black]{hyperref}
\def\TitleOfPaper{Ensemble time-dependent density functional theory}
\def\preprintlinklocation{http://dft.uci.edu}

\input Burke_template
\input macros
\usepackage{booktabs}
\usepackage{graphicx} 
\usepackage{comment}
\usepackage{caption}
\usepackage{mwe}
\usepackage{textgreek}
\graphicspath{{figures/}}

\newcommand{\kim}[1]{\textcolor{black}{#1}}

\begin{document}
\sf
\coloredtitle{\TitleOfPaper}

\coloredauthor{Kimberly J. Daas}
\email{kdaas@uci.edu}
\affiliation{Department of Chemistry, University of California, Irvine, CA 92697, USA}

\coloredauthor{Steven Crisostomo}
\affiliation{Department of Physics and Astronomy, University of California, Irvine, CA 92697, USA}

\coloredauthor{Kieron Burke}
\affiliation{Department of Chemistry, University of California, Irvine, CA 92697, USA}
\affiliation{Department of Physics and Astronomy, University of California, Irvine, CA 92697, USA}
\date{\today}

\begin{abstract}
Time-dependent density functional theory (TDDFT) is a standard approach for calculating optical excitations of molecules and solids, while ensemble DFT (EDFT) is a promising  alternative under development. We  introduce ensemble TDDFT (ETDDFT), a practical theory that combines the two, generalizing both;  we ensemble-generalize the Gross-Kohn equation and the exchange-correlation kernel of TDDFT, and generalize EDFT to time-dependent problems. We relate coordinate scaling to the adiabatic connection.  The new theory provides multiple avenues for constructing and using approximations.  We illustrate these on the 2-site Hubbard model.   We connect our results to the more general case of non-perturbative time-dependence.
\end{abstract}
\maketitle

Ground-state density functional theory (GS-DFT) in its Kohn-Sham formulation~\cite{HK64,KS65} has been remarkably successful in both materials and chemistry~\cite{B12,B14,J15,Jain2016, M20}.  But, by construction, it is designed to yield only ground-state energies and densities and possibly any excitation energy that can deduced therefrom.
Time-dependent density functional theory (TDDFT) is the standard generalization to time-dependent fields~\cite{Ullrich2011-lr,Marques2012-bn,VANLEEUWEN2001}.  The most common application is in the linear optical response regime, yielding the absorption spectrum (both transition frequencies and oscillator strengths)~\cite{c95,CASIDA1996391,BAUERNSCHMITT1996454,Grabo2000}. While successful for many routine applications, such as low-lying excitations of large molecules, its limitations with standard approximations are well documented~\cite{Maitra2022,Lacombe2023}. Real-time TDDFT yields the same results~\cite{Yabana2006}, but can also handle non-perturbative time-dependent external fields.~\cite{Kononov2022}


A promising alternative to TDDFT for extracting excitations is provided by ensemble DFT~\cite{Nagy1996,Andrejkovics1998,Nagy1995,Cernatic2021,Filatov2014,Gross1988-2,Gross1988-2,Oliveira1988,Gould2020,Gidopoulos2002}.
This theory is formulated in close analogy with ground-state DFT~\cite{Theophilou1979,Gross1988,Gross1988-2,Oliveira1988}. The HK theorem~\cite{HK64} was generalized to weighted ensembles of low-lying states, and corresponding Kohn-Sham equations for the ensemble energy and density can be defined~\cite{Gross1988,Gross1988-2,Gould2021}  and solved with some approximate ensemble (exchange correlation) XC energy~\cite{Oliveira1988}. There has been a recent explosion of interest in finding usefully accurate approximations using EDFT~\cite{Gould2017,Gould2019,Gould2020,Fromager2020}, which is now available in some quantum chemical codes~\cite{Gould2025,OCNRAK25,GKP25}, but such calculations have not yet become widespread.
While EDFT can overcome some limitations of standard TDDFT, it has some of its own, such as not predicting oscillator strengths.

\begin{figure}[ht!]
    \includegraphics[width=0.4\textwidth]{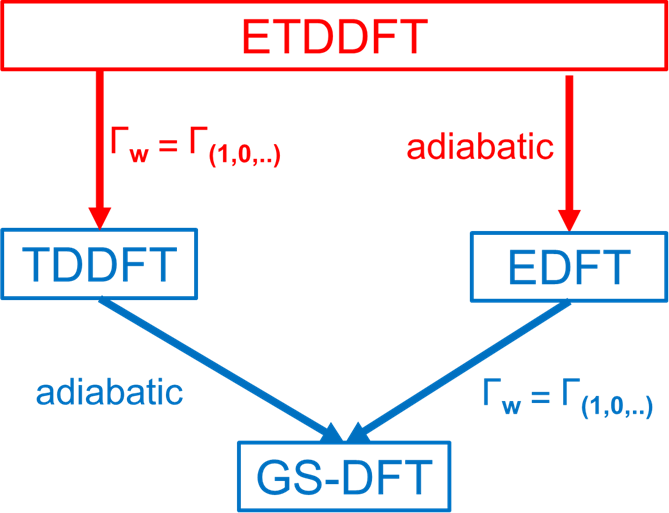}
    \label{fig:cart}
\caption{
\kim{ Logical relation between versions of DFT.}
}
\label{fig:summ}
\end{figure}

The current work generalizes the proof of TDDFT~\cite{Runge1984} to initial states that begin from an ensemble, rather than from a non-degenerate ground state.  This provides a more general theory than either TDDFT or EDFT, both of which are special cases \kim{ of ETDDFT, whose} logical relation to other forms of DFT is shown in Fig 1.  \kim{ It can also}
be considered either as a generalization of linear-response TDDFT to initial ensembles instead of pure non-degenerate ground states or a generalization of (static) ensemble DFT to time-dependent response but with time-independent weights. The one-body potential is allowed to vary in time, but the weights remain fixed. To establish its validity, we first generalize the linear response proof of van Leeuwen for the ground state \kim{\cite{VANLEEUWEN2001,Giesbertz2016}} using modified techniques developed by Pribram-Jones et al.~\cite{PriGraBur-PRL-16} for thermal DFT. The standard theorems of EDFT apply only when weights are non-increasing with excitation level, and our proof fails explicitly if these conditions are violated. The linear response proof immediately allows the generalization of the XC kernel of TDDFT to an ensemble XC kernel, which appears in a generalized Gross-Kohn formula.  We establish the connection between coordinate scaling and the coupling constant, reproducing the adiabatic-connection formula for the ensemble energy~\cite{Gould2020-2}.  We also derive various properties of the XC kernel and suggest several approximations.  All results are illustrated on the two-site Hubbard model.
We end by discussing the more general case of arbitrarily strong time-dependent fields~\cite{Runge1984}, and explain how our work is an application of an early generalization~\cite{Li1985,Li1985-2} of the RG proof to initial-state ensembles.

\noindent{\bf Background and notation:} 
TDDFT in general allows inexpensive simulation of electronic systems (both molecules and materials) in time-dependent external fields~\cite{Marques2012-bn,Ullrich2011-lr,HPB99,CarFerMaiBur-EPJB-18}. In linear response (LR)~\cite{Maitra2022}, weak electric fields are a small perturbation on the ground state KS potential, creating a proportionate time-dependent density.  The central quantity is the density-density response function for the ground state $m=0$. For the $m$th state:
\begin{equation}\label{eq:gschi}
\chi_m(\mathbf{r},\mathbf{r}',\omega) = \sum_{k\neq m} \frac{n^{*}_{mk}(\mathbf{r})n_{mk}(\mathbf{r'})}{\omega - \Omega_{km}+i\eta} - \text{c.c.}
\end{equation}
with $n_{mk}(\mathbf{r}) = \langle m | \hat{n}(\mathbf{r}) | k \rangle$ and $\hat{n}(\mathbf{r})$ being the density operator, and $\Omega_{km}$ the transition frequencies between states $k$ and $m$. The sums are over many-body states of the system.  This has a KS analog, i.e., the response function of the ground-state KS electrons in their single Slater determinant.  The full many-body $\chi$ can be found from its KS analog via the Gross-Kohn equation and the exact time-dependent XC kernel~\cite{Gross1985}.

In practice, most quantum chemical codes rewrite this expression as a matrix equation~\cite{c95,CASIDA1996391}, which has become standard.
Almost all TDDFT calculations use the {\em adiabatic} approximation~\cite{BAUERNSCHMITT1996454,Grabo2000}, ignoring the time dependence of the XC kernel, which is then just the second functional derivative of the GS XC energy.   
Even without this approximation, this scheme does not predict all desired properties~\cite{B05,B12}. In particular, the transition matrix elements between excited states are not accessed~\cite{DarMai-JCP-23,DarRoyMai-JPCL-23,DarBarMai-PRL-24}.
These are needed for \kim{ beyond-LRTDDFT (linear response TDDFT)} treatments, but require higher-order response properties~\cite{DarBarMai-PRL-24,Dar2023}.

In an unrelated theoretical development, the ensemble DFT (EDFT) also predicts excitation energies~\cite{Gross1988,Gross1988-2,Oliveira1988}, but using a variational principle analogous to the ground state case. Consider an ensemble of the ground and the first $M$ excited states ($m$), with density matrix,
\begin{equation}\label{eq:dnw}
\Gamma_{\mathbf{w}}=\sum^M_{m=0} w_{m}\ket{\Psi_m}\bra{\Psi_m}, \quad \mathbf{w}=(w_1,\ldots,w_M),
\end{equation}
whose weights ($w_m$) are monotonically non-increasing and normalized ($\sum^M_{m=0} w_m=1$). This satisfies Hohenberg-Kohn theorems of one-to-one correspondence between potentials and ensemble densities ($n_\bfw(\mathbf{r}) = \sum_{m=0}^{M} w_m n_m(\mathbf{r})$ with excited-state density $n_{m}(\br)$). 
A constrained search gives the energy as a density functional:
\begin{equation}
E_\bfw = \min_{n} \left\{ F_\bfw[n] + \int d^3r \, n(\mathbf{r}) \, v(\mathbf{r})  \right\}
\end{equation}
with $F_\bfw[n] = \min_{\Gamma_\bfw \rightarrow n} \textrm{Tr} \{ \hat{\Gamma}_\bfw (\hat{T} + \hat{V}_{\rm ee}) \}$ and $n_\mathbf{w}(\mathbf{r})$ being the minimizing density.  One constructs a non-interacting KS system with the same density and weights and defines a corresponding ensemble-dependent XC functional,
\begin{equation}\label{eq:KS}
    E_\bfw=\min_{n}\left\{T\sw[n]+\int d\mathbf{r}\, n(\mathbf{r})\,v(\mathbf{r})+E\Hxcw[n] \right\}.
\end{equation}
Here the Hartree energy has been folded in with the XC contribution, as the separation of Hartree and exchange is subtle in EDFT~\cite{Gould2017}. The user chooses how many states to include and the weights. Unlike thermal DFT, the weights of the KS systems are identical to the true systems, by construction. Transition frequencies are deduced from one or more ensemble calculations. Not only can one use ensemble energy values to extract the transition frequencies~\cite{YanPriBurUll-PRL-17}, one is also able to obtain the double transitions that adiabatic KS-TDDFT is unable to approximate~\cite{SagBur-JCP-18}. There are many excellent suggested approximations~\cite{YanPriBurUll-PRL-17,Gould2017,Gould2019,Gould2020,Gould2020-2,Gould2021}, which can overcome some limitations of standard TDDFT approaches. Recently, focus has shifted towards state specific EDFT~\cite{Gould2025} to study individual excited-states, including an investigation into static linear response EDFT~\cite{Fromager2025,DF25}. This shows the need for a time dependent extension to EDFT.


A downside of EDFT is that there are many choices of weights and any $\mathbf{w}$-dependent HXC approximation is likely to yield weight-dependent transition frequencies. The most common choices of ensemble are the original Gross-Olivera-Kohn definition (GOK) \cite{Gross1988-2}, where all weights are the same except for the highest, or GOKII, where only the ground-state has a different weight~\cite{Gross1988-2,YPBU17}.

\noindent{\bf Fundamental proof:} Assume a system begins ($t=0$) in a valid ensemble, and is weakly perturbed by $\delta v(\mathbf{r},t)$, then
\ben \label{eq:nwxw}
\delta n_\bfw(\br,t) = \int d^{3}r'\int dt' \, \chi_\bfw(\br,\br',t-t')\, \delta v(\br',t')
\een
defines the ensemble linear-response $\chi_\bfw(\br,\br',t-t')$ generating the density change $\delta n_\bfw(\br,t)$. Taking Laplace transforms with
time-coordinate $s$:
\begin{equation}
\delta n_{\bfw}(\mathbf{r},s) = \int d^3r' \, \Tilde{\chi}_\bfw (\mathbf{r},\mathbf{r}',s) \, \delta v (\mathbf{r}',s),
\label{eq:densresp}
\end{equation}
where $\Tilde{\chi}_{m}(\br,\br',s) = \chi_{m}(\br,\br',-is) $, from Eq.~(\ref{eq:gschi}). The transformed one-body operator
\begin{equation}
\delta \hat{V} (s) = \int d^3 r \,\hat{n}(\mathbf{r})\,\delta v(\mathbf{r},s)
\end{equation}
has matrix elements 
$\delta V_{ij} (s) = \langle \Psi_i \vert \delta \hat{V}(s)\vert \Psi_j\rangle$.
Consider
\begin{equation}\label{eq:Mwdn}
Y_\bfw (s) = \int d^3 r\, \delta n_\bfw (\mathbf{r},s)\, \delta v (\mathbf{r},s),
\end{equation}
where simple manipulations, see Sec.~S1 of the Supplementary Information, yield
\begin{equation}
Y_\bfw (s) = -2 \sum_{i=0}^M \sum_{j=i +1}^\infty \frac{(w_i - w_j) \Omega_{ji}}{s^2 + \Omega_{ji}^2} | \delta V_{ij} (s) |^2.
\end{equation}
As $w_i>w_j$ and  $\Omega_{ji}>0$ (assuming no degeneracies), $Y_\bfw (s)$ vanishes only if every $\delta V_{ij} (s)$ in the sum does. \kim{ Therefore, we add} a small but infinite set of weights to states above $M$, and take their weights to zero at the end of the calculation.  Then $Y_\bfw (s)$ = 0 only if $\delta v(\br,s)$ is uniform, QED.
This proof generalizes Ref.~\cite{PriGraBur-PRL-16} to any non-increasing weights  and that of Ref.~\cite{vanLeeuwen1999} to $\bfw\neq0$. 

\kim{ 
Degeneracies} couple subspaces to each other, resulting in $\Omega_{km}=0$ and a non-trivial zero density response~\cite{Giesbertz2015,PriGraBur-PRL-16}. As in Ref.~\cite{PriGraBur-PRL-16}, we assume at least $Q$ points outside the nodal hypersurface of the degenerate subspace within the $3N$ dimensional general space, where $Q$ is the number of degeneracies, ensuring that $\delta v(\mathbf{r},s)$ does not depend on $\mathbf{r}$. {\color{black} Degeneracies also limit which variations are linear, so that $\chi_{\textbf{w}}$ is not a proper functional derivative for all perturbations. This is a general issue even in standard LRTDDFT. Nevertheless, Eq.~\eqref{eq:nwxw} remains valid for those variations for which linear response is well defined~\cite{vanLeeuwen2003}.}

\noindent{\bf Formalism:} From now on, we will switch towards the more common Fourier-transformed version instead of Laplace transformed one, because in practice the differences are not relevant. However, the proof holds only for Laplace transformable potentials. Explicitly
\begin{align}\label{eq:chigen}
    \chi_{\bfw}(\mathbf{r},\mathbf{r}',\omega)=\sum^{M}_{m=0}w_m\, \chi_m(\mathbf{r},\mathbf{r}',\omega),
\end{align}
is the ensemble version of the density response function from Eq.~\eqref{eq:gschi}. Its KS analog has the same form, but its many-body states are KS Slater determinants of KS orbitals of the ground-state KS potential. Since both are invertible from our proof, we can define 
\begin{equation}\label{eq:fhxcw}
    f{\Hxcw}[n](\mathbf{r},\mathbf{r}',\omega)=\chi^{-1}\sw(\mathbf{r},\mathbf{r}',\omega)-\chi^{-1}_{\mathbf{w}}(\mathbf{r},\mathbf{r}',\omega)
    \end{equation}
as the ensemble HXC (eHXC) kernel, and the ensemble generalization of the celebrated Gross-Kohn relation~\cite{Gross1985} is \kim{ (often called the Dyson-like equation of TDDFT)}:
\begin{widetext}
\begin{equation}\label{eq:GK}
    \chi_\bfw(\br, \br',\omega)= \chi\sw(\br,\br',\omega)+\int d^3r_1\int d^3r_2\,\chi\sw(\br,\br_1,\omega)\,f\Hxcw[n](\br_1\br_2,\omega)\,\chi_{\bfw}(\br_2,\br',\omega).
\end{equation}
\end{widetext}
\kim{ Although similar to the usual GK relation, unlike LRTDDFT, the ensemble HXC kernel contains information about transitions between every eigenstate in the ensemble. We now study this kernel to understand how to extract information from it and explore approximations.}

\noindent{\bf Properties:}
We generalize our eHXC kernel to be coupling-constant dependent, by simply inserting a $\lambda$ in front of the electron-electron interaction, while holding the density fixed~\cite{LP85,GL76,HJ74} (Sec.~S2 of the SI). In the linear response case, we start with the scaling of the density response function from Eq~\eqref{eq:chigen},
\begin{equation}
\chi_\mathbf{w}^{\lambda}[n](\mathbf{r}, \mathbf{r}', \omega) 
= \lambda^4 \chi_\mathbf{w}[n_{t,1/\lambda}](\lambda \mathbf{r}, \lambda \mathbf{r}', \omega / \lambda^2),
\end{equation} 
where $n_{\lambda}(\mathbf{r},t)=\lambda^3 n(\lambda \mathbf{r},\lambda^2t)$ with $n_t=n(\mathbf{r},t)$ being the time dependent density, generalizing previous results\kim{~\cite{HPB99,Fetter2003-ng,Nagy1995,PriGraBur-PRL-16}}. With this and Eq.~\eqref{eq:GK}, we find\kim{~\cite{HPB99}}
\begin{equation}\label{eq:fhxw}
    f^\lambda\Hxcw[n](\br,\br,\omega)=\lambda^2 f\Hxcw[n_{1/\lambda}](\lambda\br,\lambda\br,\omega/\lambda^2).
\end{equation}
The \kim{adiabiatic connection fluctuation dissipation theorem (ACFDT)} for $\chi_\bfw$ is already known from Ref.~\cite{Gould2020},
\begin{widetext}
\begin{equation}\label{eq:acfd}
    E\cw[n]=-\frac{1}{2\pi}\int_0^1d\lambda \int d^3 r\int d^3 r' \frac{1}{\abs{\br-\br'}}\int_0^\infty d\omega\Im\left[\chi_{\bfw}^\lambda(\br,\br',\omega)-\chi_{\bfw}^0(\br,\br',\omega)\right],
\end{equation}
\end{widetext}
and yields the static ensemble $E\cw[n]$ (the pair density term from Ref.~\cite{Gould2020} is in the HX energy). This fluctuation dissipation theorem has a long useful history in ground-state DFT, culminating with the recent $\sigma$ functionals~\cite{Erhard2016,Trushin2021,Trushin2025}.


\noindent{\bf Approximation:} We use the ETDDFT formalism to create (perhaps too) many approximations for transition frequencies.  In TDDFT, the poles of $\chi$, or equivalently the zeroes of $\chi^{-1}$, yield transition frequencies~\cite{Runge1984,Gross1985,CASIDA1996391,c95}, which remains true for any valid ensemble.  Thus any approximate eHXC kernel yields approximate transitions, in exactly the same way as regular TDDFT, but for every valid ensemble.  We call these \emph{pole predictions}.

But we can also feed an eHXC kernel into \kim{ Eq.\eqref{eq:acfd}} to create a static ensemble correlation functional.  This also yields transition frequencies via the methods of EDFT, but only for those transitions extractable from the ensemble.  We call these \emph{ensemble predictions}.
With the exact kernel, both methods yield identical results.  Practical calculations use approximations, which typically yield distinct results.  An obvious exact condition is to recover the same result from both procedures, independent of $\mathbf{w}$.  

A note on terminology.  The adjective {\em pure} indicates a quantity evaluated with $\mathbf{w}=\mathbf{0}$, but there is an order of limits issue. Any finite weight, no matter how small, generates new poles in $\chi_{\mathbf{w}}$, which yield a zero in $\chi_{\mathbf{w}}^{-1}$ and so a new transition, e.g., between two excited states.  The strength of this pole vanishes linearly with $\mathbf{w}$.  Any ensemble approximation is {\em almost pure} if all its excited-state weights are taken as approaching, but not equal to, zero.  The adjective \kim{ \emph{adiabatic}} indicates a quantity evaluated at $\omega=0$.  For finite systems, static quantities are simply ground-state quantities, and there is no ambiguity. \kim{ This creates 4 combinations, pure adiabatic ($w\rightarrow0$, $\omega\rightarrow0$), ensemble adiabatic ($\omega\rightarrow0$), pure dynamic ($w\rightarrow0$) and ensemble dynamic, i.e., exact.}

To see the myriad ways ETDDFT can be used, begin with the usual approach in TDDFT, namely to use a ground-state XC functional,
producing a static approximation to $f\xc$.  Pole prediction often yields good approximations to low-lying optical excitations, but misses double excitations,  which require frequency dependent kernels\kim{~\cite{DarMai-JCP-23,CZMB04,Maitra2004,Maitra2022,Dar2025}}.  On the other hand, one can also use the adiabatic kernel in the \kim{ ACFDT} formula, \kim{ to find approximations} for the ground-state $E\c$.  With ETDDFT, because $\chi\s$ depends on $\mathbf{w}$, the pole-predictions now yield approximate transitions for every valid set of ensemble weights, reducing to the TDDFT results when $\mathbf{w}\rightarrow0$. On the other hand, the ensemble predictions from the \kim{ ACFDT} {\em also} yield transition frequencies among members of the ensemble for every set of valid weights, including transitions between unoccupied states (which can be very useful~\cite{DarRoyMai-JPCL-23,DarMai-JCP-23}, but usually require higher-order TDDFT response~\cite{DarBarMai-PRL-24,Dar2023}).

\begin{figure}[htb]
    \includegraphics[width=0.4\textwidth]{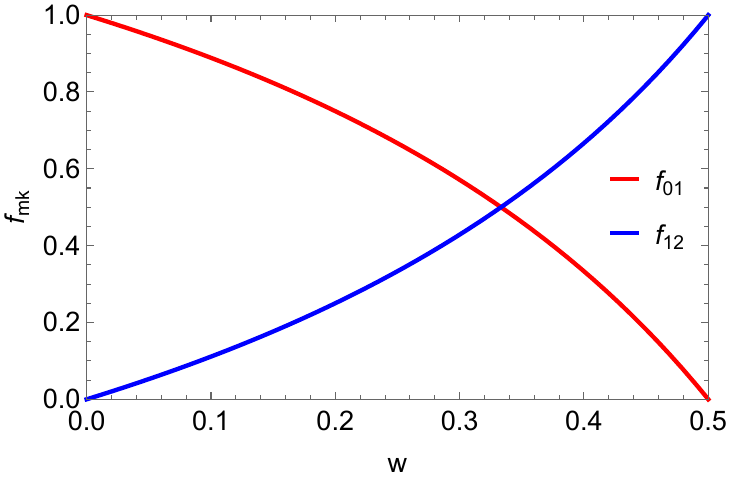}
    \label{fig:osc}
\caption{
\kim{ The weight dependence of the oscillator strengths of the first excitation $f_{01}$ and the excited state to excited state transition $f_{12}$ at $U=t$.
}}
\label{fig:osc}
\end{figure}
\noindent{\bf Illustrations}: We illustrate ETDDFT on the 2-site Hubbard model at half filling, which is useful for demonstrating basic principles of DFT~\cite{Carrascal_2015,CarFerMaiBur-EPJB-18,Deur2018,SKCPB23}, due to its tiny Hilbert space and analytic solutions.
Here we need only the symmetric case to make our points:
\begin{equation}\label{eq:HHubbard}
    \hat{H} = -t \sum_{\sigma} (\hat{c}^\dagger_{1\sigma} \hat{c}_{2\sigma} + h.c.) 
+ U \sum_{i} \hat{n}_{i\uparrow} \hat{n}_{i\downarrow}, 
\end{equation}
with $t$ the hopping parameter and $U$ the onsite repulsion.  Exact expressions \kim{ for the} asymmetric case are given in Secs.~S3-S5 of the SI.
As in Ref.~\cite{SKCPB23}, we restrict ourselves, to the three singlet states: the ground state, the \kim{ first (predominately single) excitation and second (predominately double) excitation~\cite{ZhaBur-PRA-04}
and consider}
\begin{equation}\label{eq:hubens}
\Gamma_w=\bar{w}\ket{\Psi_0}\bra{\Psi_0}+w\ket{\Psi_1}\bra{\Psi_1},
\end{equation}
with $\bar{w}=1-w$ and $0\leq w<0.5$. Because of norm conservation, the density can be characterized by a single number (usually $n_2-n_1$), so $\chi(\omega)$ is just a function.
The many-body $\chi_w(\omega)$ is a weighted combination (Eq.~\eqref{eq:chigen}) of the ground-state $\chi_{0}(\omega)$
\begin{equation}\label{eq:chi0}
    \chi_{0}(\nu)=\frac{2 \nu_1 A_{01}}{\nu_{+}^2-\nu_1^2}+\frac{2 \nu_2 A_{02}}{\nu_{+}^2-\nu_2^2}\\
\end{equation}
with $\nu_+=\omega/(2t)+i\eta$, and of the singly-excited state:
\begin{equation}\label{eq:chi1}
    \chi_{1}(\nu)=-\frac{2 \nu_1 A_{01}}{\nu_{+}^2-\nu_1^2}+\frac{2 \Delta \nu_2 A_{12}}{\nu_{+}^2-\Delta{v}^2},\\
\end{equation}
which includes the transition back to the ground state and
a transition upward to the double excitation at frequency ($\nu_2-\nu_1)$, where $A_{mk}=\left|\bra{m}\Delta \hat{n}\ket{k}\right|^2$ (see Eq.~S19). The exact eHXC kernel is given by Eq.~\eqref{eq:fhxcw}.
Some algebra yields poles at~\cite{CASIDA1996391,c95,CarFerMaiBur-EPJB-18} \kim{ (Casida equation)},
\begin{align}
\nu^2&=\nu_s^2+2\nu\s\left((1-2w) A_{{\sss{S}},1}+w\, A_{{\sss{S}},2}\right)f_{{\rm Hxc},w}(\nu),\label{eq:cas}
\end{align}
and one can also extract oscillator strengths in the usual way (Eq.~30-32 of Ref.~\cite{CarFerMaiBur-EPJB-18}), \kim{ which yields
\begin{equation}
    f_{mk,w}=g_{mk,w}(e_{m}-e_{k})A_{mk},
\end{equation} 
where $g$'s are weights and $e_m$ is the energy of the $m$th eigenstate
(see Eq.~\ref{eq:fosc} for derivation). Figure~\ref{fig:osc} shows the exact oscillator strengths, including between excitations, extracted from the ensemble kernel of Eq.~\ref{eq:fhxcw}.}

\begin{figure*}[t!]
\centering
\begin{minipage}{0.45\linewidth}
    \centering
    \includegraphics[width=\textwidth]{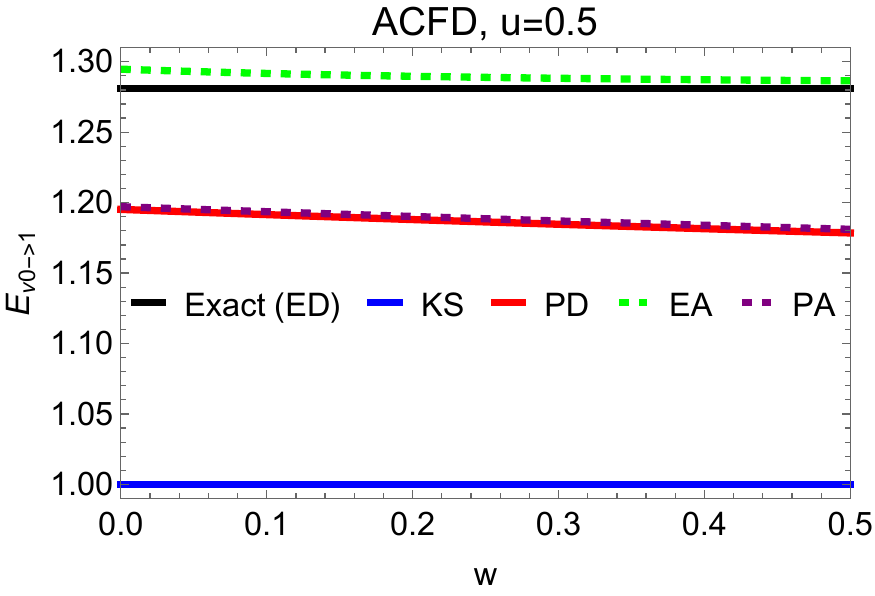}
    \label{fig:h-b2}
\end{minipage}
\begin{minipage}{0.45\linewidth}
    \centering
    \includegraphics[width=\textwidth]{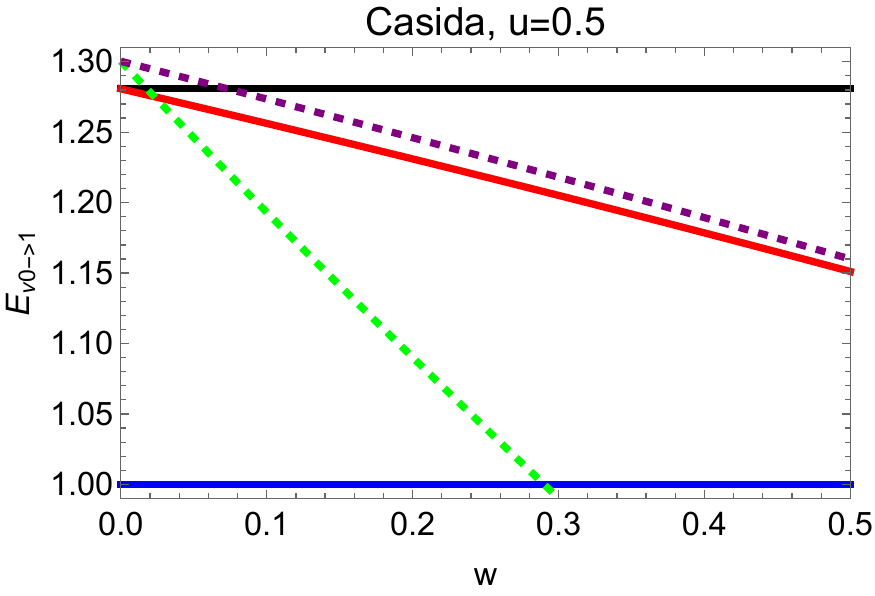}
    \label{fig:h-c2}
\end{minipage}
\caption{
\kim{ Two alternative applications of ETDDFT: the} weight dependence of the transition from the ground to first excited state for $U=t$ for the exact, KS and three approximations (pure dynamic, ensemble adiabatic, and pure adiabatic). More figures in Sec.~S6 of the SI.
}
\label{fig:approxs}
\end{figure*}

Figure~\ref{fig:approxs} shows transition frequencies at $U=0.5$ found from $(E_w-E_0)/w$ in the ensemble case, and from solving the pole equation, with various approximations.  While the exact kernel yields the exact answer for any
allowed $w$, the approximations yield transition frequencies that vary significantly with $w$.  One way out of this maze is to consider only the prediction as $w\to 0$, which can produce lots of other simplifications too~\cite{YanPriBurUll-PRL-17}.  For this simple model and ensemble, solving TDDFT equations is usually more accurate than extracting ensemble energies.  However,
for the ensemble adiabatic approximation, $f_{{\rm Hxc},w}(0)$, the ensemble prediction is competitive or sometimes better. \kim{ Figure~\ref{fig:approxs} is specific to the Hubbard dimer, and cannot be used to judge the accuracy for more realistic systems: It illustrates what is possible, not what is accurate.}


\noindent{\bf Arbitrary time-dependence:} The RG proof of uniqueness
allows us to write $v[\n,\Psi(0)](\br,t)$, i.e., for given statistics (fermions) and electron-electron interaction (Coulomb repulsion), the time-dependent one-body potential is a functional of the time-dependent density and initial wavefunction (compatible with the initial density).~\cite{Runge1984}  In turn, in the TDKS equations, this implies 
$v\xc[\n,\Psi(0),\Phi(0)](\br,t)$, where $\Phi(0)$ is the initial KS wavefunction.
For an initial non-degenerate ground-state density, the ground-state wavefunction is a functional of that density according to HK, making everything a functional of the time-dependent density alone.

Almost as soon as the RG~\cite{Runge1984} proof appeared, it was generalized to the case of an initial density matrix, i.e., not just a pure state~\cite{Li1985,Li1985-2}, yielding 
$v\xc[\n,\hat{\Gamma}(0),\hat{\Gamma}_s(0)](\br,t)$.  In fact, applying coordinate scaling to this case, yields
\begin{widetext}
\begin{equation}
    v^\lambda_{xc}[n_t,\hat{\Gamma}(0),\hat{\Gamma}_{\sss{S}}](\br,t) =\lambda^2v_{xc}[n_{t,1/\lambda},\hat{\Gamma}_{1/\lambda}(0),\hat{\Gamma}_{{\sss{S}},1/\lambda}(0)](\lambda\br,\lambda^2\t),
\end{equation}
\end{widetext}
where we used the scaling of a general density matrix as 
\begin{equation}
\hat{\Gamma}^{\lambda}[n_t,\hat{\Gamma}(0)]=\hat{\Gamma}_{\lambda}[n_{t,1/\lambda},\hat{\Gamma}_{1/\lambda}(0)],
\end{equation}
as shown in Sec.~S2.
Restricting ourselves to density matrices with non-increasing weights, we apply the GOK theorem~\cite{Gross1988-2,Gross1988,Oliveira1988} to eliminate the dependence on the initial density matrices:
\begin{equation}
   v\Hxcw[\n_t](\br,t) = v\Hxc[\n_t,\hat{\Gamma}_\mathbf{w}(0)[\n_0],\hat{\Gamma}\sw(0)[\n_0]](\br,t) 
\end{equation}
with $n_0$ being the density at $t=0$, i.e., a pure weight-dependent functional of the time-dependent density only.
Our eHXC kernel \kim{ Eq.~\eqref{eq:fhxcw}} is simply the functional derivative of this $v\Hxcw (\br,t)$.  On the other hand, our linear response proof used earlier avoids the complications of power series expansions in time~\cite{vanLeeuwen1999,VANLEEUWEN2001}.

In Ref.~\cite{Giesbertz2016} it was shown that $\chi_\bfw$ is always invertible if the weights are monotonic, depend only on the energy, and all states are included in the ensemble. The first is required~\cite{PriGraBur-PRL-16,Deur2018,SKCPB23}, and the second also holds, because the weights are simply constants. To satisfy the last, we can imagine occupying all states higher than $M$ with an infinitesimal weight, and then taking the limit. 

\noindent{\bf Summary:} In this work, \kim{ we showed that ETDDFT  has the advantages of both EDFT and TDDFT, e.g. being able to treat all excited states and having access to all oscillator strengths. Our theory defines a new functional, the ensemble HXC kernel, for which we also derived the exact coupling constant scaling relations and properties of this kernel and potential.} We have defined several classes of approximations that can be used to calculate the transition frequencies via the matrix formulation and the \kim{ ACFDT}. Indeed, we show that these approximations have widely different $\mathbf{w}$-dependence for the Hubbard dimer\kim{, see Fig.~2}.

\kim{ Our work also extends the applicability of LRTDDFT to other quantities, such as Rabi oscillations~\cite{DarBarMai-PRL-24}, which were require quadratic response in standard TDDFT. Our extension is relatively easy to implement in standard quantum chemical packages because it is based on the well established LRTDDFT language that all such codes already use~\cite{CASIDA1996391,c95,Maitra2022}.
An investigation of ETDDFT approximations for realistic systems is planned.}
 
 \noindent{\bf Acknowledgments} K.J.D. would like to thank UCI's Chancellors Postdoctoral Fellowship Program and in particular Prof. Dr. Feizal Waffarn for his support as a sponsor. S.C. acknowledges support from the UC Presidential Dissertation Year (PDY) fellowship. S.C. and K.B. were supported by the NSF Award No. CHE-2154371. We like to acknowledge Klaas Giesbertz for fruitful discussions.

\newpage

\begin{center} \noindent{\bf End Matter}\end{center}

 \noindent{\bf Oscillator Strengths in the Hubbard Dimer}

In this section we generalize the work of Ref.~\cite{CarFerMaiBur-EPJB-18} for ensembles. From Eqs.~29 and 30 of Ref.~\cite{CarFerMaiBur-EPJB-18}, after some simple algebra we find,
\begin{equation}
    {G_{i,w}} = \sqrt{\frac{m_i(w)\nu_i A_i}{(1-2w)\nu_s A_{s,1} + w\Delta v_s A_{s,2}}},
\end{equation}
where $i\in[01,02,12]$ and $m_i(w)=\{1-2w,1-w,w\}$. With this the oscillator strengths are given by
\begin{equation}\label{eq:fosc}
f_{i,w}=\frac{G_{i,w}^2}{\sum_j G_{j,w}^2} = m_i(w)\nu_i A_i/v_{3,w},
\end{equation}
where $\nu_{3,w}={\sum_jm_j(w)v_jA_j}$ is chosen so  $\sum_i f_{i,w}=1$. With this auxiliary quantity, we can rewrite the $\chi_w(\omega)$ as,
\begin{equation}
\chi_w(\omega)=2\nu_{3,w}\left(\frac{f_{1,w}}{\nu_{+}^2-\nu_1^2}+\frac{f_{2,w}}{\nu_{+}^2-\nu_2^2}+\frac{1-f_{1,w}-f_{2,w}}{\nu_{+}^2-\Delta{\nu}^2}\right).
\end{equation}

\noindent{\bf Small Matrix and Single Pole Approximations}

The small matrix approximation (SMA) and single pole approximation (SPA) can be derived in a similar fashion as the general Casida equation~\cite{CarFerMaiBur-EPJB-18}. The only difference is that we assume that the poles of $\chi_w(\omega)$ are well separated. To derive the SMA, we assume that instead of looking at all the poles at once, we only look at transitions near a certain many body transition, $\Omega_{i}\rightarrow\nu_{i}$, which means that only the pole near this transition frequency contributes. After some algebra, we find
\begin{equation}
\nu_{i}^2=\nu_{{\sss{S}},i}^2+2m_{i}(w)\nu_{{\sss{S}},i}A_{{\sss{S}},i}f_{\rm HXC,w}(\nu_{{\sss{S}},i}),
\end{equation}
with $v_{{\sss S},1}=v_{\sss{S}}$ and $v_{{\sss S},2}=\Delta v_{\sss{S}}$. We can find the SPA by assuming that there is only a small shift from the KS value, which gives,
\begin{equation}
    \nu_{i}=\nu_{{\sss{S}},i}+{2m_{i}(w)}f_{\rm Hxc,w}(v_{{\sss{S}},i}).
\end{equation}
\bibliography{steven}

\label{page:end}
\end{document}

%% file: Burke_template.tex
\usepackage{amsmath}
\usepackage{physics}
\usepackage{amssymb}
\usepackage{float}
\usepackage[dvipsnames]{xcolor}
\usepackage{subcaption}
\usepackage{multirow}
\usepackage{caption}

\definecolor{TITLECOL}{rgb}{0.05,0.25,0.85}
\definecolor{CONTENTSCOL}{rgb}{0.1,0.2,0.7}
\definecolor{URLCOL}{rgb}{0,0.52,0.83}
\definecolor{LINKCOL}{rgb}{0.05,0.5,0}
\definecolor{CITECOL}{rgb}{0.25,0,0.48}
\definecolor{SECOL}{rgb}{0.07,0.31,0.80}
\definecolor{SSECOL}{rgb}{0.26,0.19,0.75}

\newcommand{\coloredtitle}[1]{\title{\textcolor{TITLECOL}{#1}}}
\newcommand{\coloredauthor}[1]{\author{\textcolor{CITECOL}{#1}}} 

\usepackage{graphicx}

\def\PublicationsLink{http://dft.uci.edu/publications.php}
\def\preprintlink{ \href{\preprintlinklocation}{\TitleOfPaper} }
\def\preprinttext{~}

\usepackage{fancyhdr}
\makeatletter
\fancypagestyle{titlepage}
{	\lhead{\textsc{\preprinttext\ ~ \href{\PublicationsLink}{~}}}
	\chead{}
	\rhead{\preprintlink}
	\lfoot{}}
\makeatother
\def\preprintlink{ 
	\href{\preprintlinklocation}
        {
~}
	}

\definecolor{Green}{rgb}{0.016,0.627,0}
\definecolor{Plum}{rgb}{0.17,0,0.45}
\definecolor{LBlue}{rgb}{0,0.34,0.45}
\definecolor{Sepia}{rgb}{0.37,0.17,0.02}
\definecolor{BurntOrange}{rgb}{0.78,0.39,0}

\usepackage{hyperref}
\hypersetup{
    breaklinks=true,
    bookmarksopen=true,
    bookmarksnumbered=true,
    colorlinks=true,
    linkcolor=LINKCOL,
    linktocpage=true,
    citecolor=CITECOL,
    filecolor=magenta,      
    urlcolor=URLCOL,
}

%% file: macros.tex
\def\bea{\begin{eqnarray}}
\def\eea{\end{eqnarray}}
\def\ben{\begin{equation}}
\def\een{\end{equation}}
\def\benu{\begin{enumerate}}
\def\enu{\end{enumerate}}

\def\bei{\begin{itemize}}
\def\eei{\end{itemize}}
\def\beit{\begin{itemize}}
\def\eit{\end{itemize}}
\def\benu{\begin{enumerate}}
\def\enu{\end{enumerate}}

\def\n{n}

\def\sss{\scriptscriptstyle\rm}

\def\1var{(\bx_1...\bx\N)}

\def\br{{\bf r}}

\def\bx{{\bf x}}

\def\bfw{{\mathbf{w}}}

\def\c{_{\sss C}}
\def\s{_{\sss S}}

\def\xc{_{\sss XC}}
\def\cw{_{\sss c,\mathbf{w}}}

\def\Hxcw{_{{\sss HXC},\mathbf{w}}}

\def\Hxc{_{\sss HXC}}

\def\N{_{\sss N}}

\def\sph_int{ {\int d^3 r}}

\def\t{t}

\def\sw{_{{\sss S},\mathbf{w}}}

\def\w{_{w}}